\documentclass[conference]{ieeeconf}
\IEEEoverridecommandlockouts
\usepackage{cite}
\usepackage{amsmath,amssymb,amsfonts}
\usepackage{algorithmic}
\usepackage{graphicx}
\usepackage{multirow}
\usepackage{textcomp}
\usepackage{booktabs}
\usepackage{xcolor}
\usepackage{float}
\usepackage[symbol]{footmisc}
\usepackage{wrapfig,lipsum,booktabs}
\usepackage{hyperref}

\def\BibTeX{{\rm B\kern-.05em{\sc i\kern-.025em b}\kern-.08em
    T\kern-.1667em\lower.7ex\hbox{E}\kern-.125emX}}
\begin{document}

\title{Learning Scalable Policies over Graphs for Multi-Robot Task Allocation using Capsule Attention Networks}



\author{Steve Paul$^{1}$, Payam Ghassemi$^{2}$, and Souma Chowdhury$^{3,\dagger}$
\thanks{Authors $^{1}, ^{2}, ^{3}$ are with the department of Mechanical and Aerospace Engineering,
        University at Buffalo, Buffalo, NY, USA
        {\tt\small \{stevepau, payamgha, soumacho\}@buffalo.edu}}%
\thanks{$^\dagger$ Corresponding Author, soumacho@buffalo.edu}
\thanks{This work was supported by the Office of Naval Research (ONR) award N00014-21-1-2530 and National Science Foundation (NSF) award CMMI 2048020. Any opinions, findings, conclusions, or recommendations expressed in this paper are those of the authors and do not necessarily reflect the views of the ONR or the NSF.}
\thanks{Copyright\textcopyright2022 IEEE. Personal use of this material is permitted. Permission from IEEE must be obtained for all other uses, in any current or future media, including reprinting/republishing this material for advertising or promotional purposes, creating new collective works, for resale or redistribution to servers or lists, or reuse of any copyrighted component of this work in other works.}
}

\maketitle

\begin{abstract}
This paper presents a novel graph reinforcement learning (RL) architecture to solve multi-robot task allocation (MRTA) problems that involve tasks with deadlines and workload, and robot constraints such as work capacity. While drawing motivation from recent graph learning methods that learn to solve combinatorial optimization (CO) problems such as multi-Traveling Salesman and Vehicle Routing Problems using RL, this paper seeks to provide better performance (compared to non-learning methods) and important scalability (compared to existing learning architectures) for the stated class of MRTA problems. The proposed neural architecture, called Capsule  Attention-based Mechanism or CapAM acts as the policy network, and includes three main components: 1) an encoder: a Capsule Network based node embedding model to represent each task as a learnable feature vector; 2) a decoder: an attention-based model to facilitate a sequential output; and 3) context: that encodes the states of the mission and the robots. To train the CapAM model, the policy-gradient method based on REINFORCE is used. When evaluated over unseen scenarios, CapAM demonstrates better task completion performance and $>$10 times faster decision-making compared to standard non-learning based online MRTA methods. CapAM's advantage in generalizability, and scalability to test problems of size larger than those used in training, are also successfully demonstrated in comparison to a popular approach for learning to solve CO problems, namely the purely attention mechanism. 
\end{abstract}



\section{Introduction} 
\label{sec:Introduction}


Multi-Robot Task Allocation (MRTA) involves coordinating a set of tasks among a team of cooperative robotic systems such that the decisions are free of conflict and optimize a quantity of interest~\cite{gerkey2004formal}. The potential real-world applications of such MRTA problems are immense considering that multi-robotics is one of key emerging directions of robotics research and development, and task allocation is fundamental to most multi-robotic or swarm-robotic operations. Examples include disaster response~\cite{ghassemi2018decmata}, last-mile delivery~\cite{aurambout2019last}, environment monitoring~\cite{espina2011multi}, and reconnaissance~\cite{olson2012progress}.
We specifically focus on a class of MRTA problems that falls into the Single-task Robots, Single-robot Tasks, and Time-extended Assignment (SR-ST-TA) class defined in \cite{gerkey2004formal,nunes2017taxonomy}, that include characteristics such as tasks with time deadlines and workload, and robots with constrained work capacity. Based on iTax taxonomy as defined in \cite{gerkey2004formal}, these problems fall into the In-schedule Dependencies (ID) category. 
The MRTA problems considered here involves \textit{mildly-heterogeneous} robots (with different task completion rates), and where the task selections are assumed to take place in a decentralized manner, under full observability of the task environment and peers' decisions.

There exists a notable body of work (mostly non-learning approaches) in solving MRTA problems related to these applications, e.g., graph-based methods~\cite{ghassemi2018decmata}, integer-linear programming (ILP) approaches~\cite{nallusamy2009optimization,toth2014vehicle}, and auction-based methods~\cite{dias2006market}. However, critical challenges continue to persist, especially regarding the scalability with number of robots and tasks, and the ability to adapt to complex problem characteristics without tedious hand-crafting of underlying heuristics. 
Existing solutions to multi-Traveling Salesman Problem (mTSP) and Vehicle Routing Problem (VRP) in the literature~\cite{bektas2006multiple,braekers2016vehicle} have addressed analogical problem characteristics of interest to MRTA, albeit in a disparate manner; these characteristics include tasks with time deadlines, and multiple tours per vehicle, with applications in the operations research and logistics communities~\cite{azi2010exact,wang2018multi}.
Integer linear programming or ILP based mTSP-type formulations and solution methods have also been extended to task allocation problems in the multi-robotics domain~\cite{jose2016task}. Although the ILP-based approaches can in theory provide optimal solutions, they are characterized by exploding computational effort as the number of robots and tasks increases~\cite{toth2014vehicle,cattaruzza2016vehicle}.
For example, for the studied SR-ST problem, the cost of solving the exact ILP formulation of the problem, even with a linear cost function (thus an ILP), scales with $\mathcal{O}(n^3m^2h^2)$, where $n$, $m$, and $h$ represent the number of tasks, the number of robots, and the maximum number of tours per robot, respectively~\cite{ghassemi2019decmrta}.
As a result, most online MRTA methods, e.g., auction-based methods~\cite{dias2006market,schneider2015auction,otte2020auctions}, metaheuristic methods \cite{Mitiche2019, VANSTEENWEGEN20093281}, and bi-graph matching methods~\cite{ismail2017decentralized,ghassemi2019decmrta,Ghassemi2020}, use some sort of heuristics, and often report the optimality gap at least for smaller test cases compared to the exact ILP solutions.


In recent years, learning approaches based on Graph Neural Networks or GNN (which can model non-Euclidean data) are being increasingly used to solve planning problems with a Combinatorial Optimization (CO) formulation, e.g., TSP, VRP, Max-Cut, Min-Vertex, and MRTA~\cite{Kool2019, barrett2019exploratory, khalil2017learning, Kaempfer2018LearningTM, mittal2019learning, li2018combinatorial, nowak2017note, 9116987, Tolstaya2020MultiRobotCA, Sykora2020, Dai2017}. 
These existing studies are however limited in three key aspects: \textbf{1)} They address simplified problems that often exclude common real-world factors such as resource and capacity constraints \cite{Kool2019, Kaempfer2018LearningTM, khalil2017learning, Tolstaya2020MultiRobotCA}). \textbf{2)} They are mostly focused on smaller sized problems ($\leq$ 100 tasks)~\cite{Paleja2020,Malcolm2005, 9116987,Sykora2020}, with their scalability remaining unclear. \textbf{3)} They rarely provide evidence of generalizing to problem scenarios that are larger in size than those used for training. This latter capability is particularly critical since real-world MRTA problems often involve simulating episodes whose costs scale with the number of tasks and robots, making re-training efforts burdensome. Moreover the time-sensitive nature of many real-world applications of MRTA \cite{lin2018efficient}, such as in disaster response \cite{Ghassemi2020}, call for a near real-time performance, while involving a large-sized decision space \cite{9435208}. Therefore, near real-time computable good feasible solutions are of essence.

Along with the need to address scalability, we posit that a sequential decision-making approach is favored considering environmental and communication uncertainties and the possibility of new tasks to emerge during the operation \cite{8206001, 9272626}. Therefore, we formulate the MRTA problem such that the decision taken by the robots on visiting the next node is performed in a sequential, decentralized, and asynchronous manner.
To enable scalable policies, we design a  novel \textbf{encoder-decoder} policy network, where the \textbf{encoder} is based on Graph Capsule Convolutional Neural Networks (GCAPCN) \cite{Verma2018}, which is hypothesized to incorporate local and global structural information with permutation invariance. The \textbf{decoder} is based on a Multi-head Attention mechanism (MHA)~\cite{Kool2019, VaswaniSPUJGKP17} which fuses the encoded information and problem-specific \textbf{context} using matrix multiplication, in order to enable decentralized sequential decisions. The proposed network architecture is named Capsule Attention Mechanism or CapAM. 

\textbf{Contributions:} 
The main contributions of this paper can be summarized as \textbf{1)} Formulate the general SR-ST class of MRTA problems as a Markov Decision Process or MDP over graphs with the multi-robot system's state embedded as the \textit{context} portion of the policy model, such that the task allocation policy can be learned using an RL approach. \textbf{2)} Explore how a policy network that learns the structural information of task-graphs results in better decision-making. \textbf{3)} Demonstrate this learning framework's ability to generalize to larger-sized problems without the need to retrain. \vspace{-.1cm}

\textbf{Paper outline: }
The next section provides a brief overview of MRTA and it's MDP formulation. Section \ref{sec:Learning_framework} then describes our proposed new graph learning architecture that operates over this MDP. Section \ref{sec:Experimental_evaluation} presents the settings and outcomes of numerical experiments on a large number of task allocation problems with time and capacity constraints \cite{Mitiche2019}, which are used to evaluate this new approach to MRTA. This evaluation study includes comparative analysis w.r.t. non-learning methods as well as a well-known attention mechanism based reinforcement learning method, and further parametric and scalability analysis. Finally, Section \ref{sec:Conclusion} provides concluding remarks and discusses the potential future extensions for our new graph learning-based approach to MRTA.
Throughout this paper, \textbf{tasks} are also referred by the terms ``\textbf{nodes}'' or ``\textbf{vertices}''.
\section{MRTA: Problem definition and formulation}
\label{sec:problem_formulation}

MRTA problems can be formulated as Integer Linear or Non-Linear Programming  (ILP or INLP) problems. When tasks are defined in terms of location, the MRTA problem becomes analogical to the Multi-Traveling Salesmen Problem (mTSP)~\cite{khamis2015multi} and its generalized version, the Vehicle Route Planning (VRP) problem~\cite{dantzig1959truck}, albeit with additional constraints and operation-specific objective function. \vspace{-.2cm}
\subsection{MRTA problem description and formulation}
\label{sec:mrta_optimization}

The exact solution of the MRTA problem can be obtained by formulating it as the following INLP problem:

\vspace{-.2cm}
\begin{align}
    \label{eq:objectiveFunction}
 \min~ f_\text{cost} =  \sum_{i = 1}^{N} r_{i}, \
\begin{cases}
    r_{i} = \frac{t_{i}^{f}}{d_{i}},& \text{if } \ t_{i}^{f} > d_{i}\\
    0,              & \text{otherwise}
\end{cases}
\end{align}
 \ \ \  \ \  \text{subject to~} 
\begin{align}
\label{eq:constraint1}
& s_{i} \in S \; \forall \; i \in [1..N] \\
\label{eq:constraint2}
& s_{i} \neq s_{j} \; \forall \ i \neq j
\end{align}

\noindent Here $t_{i}^{f}$ is the time at which task $i$ was completed, $d_{i}$ is the time deadline of task $i$, and $S = [s_{1}, s_{2}, ... s_{N}]$ is the sequence of all the tasks/nodes that were visited, and $N$ is the number of tasks available. The minimum cost function that can be achieved using Eq. \ref{eq:objectiveFunction} is $0$, which corresponds to the case where all the tasks are successfully completed.
A detailed formulation of the exact ILP constraints that describe the MRTA problem can be found in this recent work on MRTA ~\cite{Mitiche2019}. Note that, in our paper, we use a slightly different objective/cost function as compared to \cite{Mitiche2019}, but the intention of the objective function of both of these works remains the same, which is to maximize the number of successfully completed tasks (thereby allowing fair comparison with methods reported in the literature). 
Here, we craft the objective function (Eq.~\eqref{eq:objectiveFunction}) such that only missed tasks ($t_{i}^{f} > d_{i}$) contribute to the cost function. This objective function can be tailored according to user/application preferences, without affecting the proposed learning framework. 
The constraints in Eqs. \ref{eq:constraint1} and \ref{eq:constraint2} are such that each task must be visited exactly once by any robot.

\begin{figure}[!t]
    \centering
    \includegraphics[width=.9\linewidth]{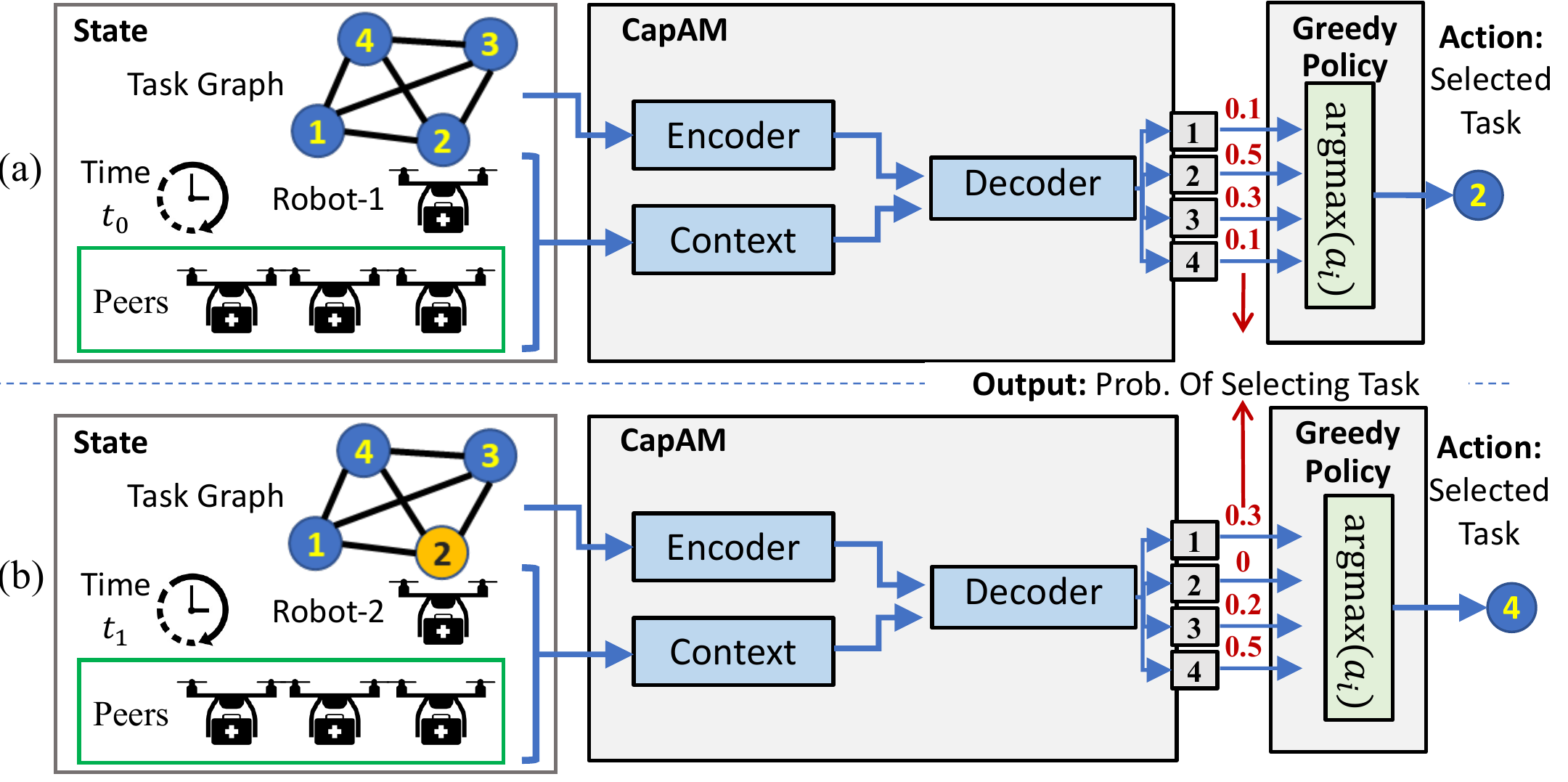}
    \vspace{-0.4cm}
    \caption{Deployment of an MRTA policy based on the CapAM architecture. a) Robot-1 at $t_0$. b) Robot-2 at $t_1$; here, the output for previously selected task (e.g., task 2 in (b)) is set at 0.}
    \label{fig:sequential_decision_making}
    \vspace{-0.6cm}
\end{figure}

\subsection{An ``MDP over Graph" Representation of MRTA}\label{sec:ProblemFormulation}
The task space of an MRTA problem can be represented as a complete un-directed graph $\mathcal{G} = (V, E, \textbf{A})$, which contains a set of nodes/vertices ($V$), a set of edges ($E$) that connect the vertices to each other, and the weighted adjacency matrix $\textbf{A}$ that gives the extent to which two nodes are connected. Each node is a task, and each edge connects a pair of nodes. For MRTA with $N$ tasks, the number of vertices and the number of edges are $N$ and $N(N-1)/2$, respectively. Node $i$ is assigned a 3-dimensional feature vector denoting the task location and time deadline, i.e., $\delta_i=[x_i,y_i,d_i]$. Here we consider a weighted adjacency matrix without self-loop ($\alpha_{ii} = 0, \alpha_{i,j} \in \textbf{A}$, where $ i \ {\text{and}} \ j \in [1, N]$). 
The weights in the adjacency matrix $\alpha_{ij} = 1/(1+ |\delta_i - \delta_j|)$ take a real value between 0 and 1, in a manner such that if the features of two nodes $i$ and $j$ are relatively closer to each other, then $\alpha_{ij}$ is higher. 
Since we formulate MRTA as an undirected graph, the adjacency matrix $A$ is symmetric. In this paper, the elements of $\textbf{A}$ ($\alpha_{i,j}$, $i \neq j$) are computed as the inverse of the normalized Euclidean distance of the node properties. 

The MDP can be defined in a decentralized manner to capture the myopic task-assignment process of each individual robot, which can be expressed as a tuple $<\mathcal{S}, \mathcal{A}, \mathcal{P}_a, \mathcal{R}>$. Figure \ref{fig:sequential_decision_making} shows how this MDP is executed to perform the sequential task selection by each robot. As evident from Fig. \ref{fig:sequential_decision_making}, the components of the MDP can then be defined as follows: \textbf{State Space ($\mathcal{S}$):} A robot at any decision-making instance uses a state $s\in\mathcal{S}$, which contains the following information: 
1) elapsed mission time, 2) robot's current location, 3) constraints of this robot such as its work capacity, 4) the planned tasks of its peers, and 5) the constraints of its peers. The state space also includes the environment, defined by the location and time deadline of tasks, which are expressed here as node features in the graph. Here we assume that each robot can broadcast its information, which includes the task it is currently engaged in and its work capacity, to its peers without the need for a centralized system for communication; this is well aligned with modern communication capabilities \cite{Sykora2020}.
\noindent\textbf{Action Space ($\mathcal{A}$):} The set of actions is represented as $\mathcal{A}$, where each action $a\in\mathcal{A}$ is defined as the index of the selected task, $\{1,\ldots,N\}$. In MRTA problems that include a depot, task $0$ can be considered as the depot. The task status, i.e., \textit{active}, \textit{completed}, and \textit{missed} (i.e., deadline is passed), is used to properly decode the actions generated by the policy model, with the latter being in the form of probability of selecting each available task. Note that even under full observability, as the action space of a single robot includes only its own degree of affinity for selecting each available task, the decentralized decision-making formulation differs from an ideal centralized formulation, and remains readily scalable. \textbf{Transition ($\mathcal{P}_a$):} The transition is an event-based trigger. An event is defined as the condition that a robot reaches its selected task or visits the depot. As environmental uncertainties and communication issues (thus partial observation) are not considered in this paper, only deterministic state transitions are allowed. 
\textbf{Reward ($\mathcal{R}$):} A delayed reward is considered here, estimated at the end of the multi-robot simulation, i.e,  when there are no more active tasks. The actual reward function is typically application-dependent. For example, in some MRTA problems, task completion is more important, while in another, decreasing the total effort (e.g., distance traveled) by robots is more important, which can then be estimated using the weight matrix.

\vspace{-.1cm}

\section{Learning framework} 
\label{sec:Learning_framework}
\begin{figure}[!t]
    \centering
    \includegraphics[width=0.46\textwidth]{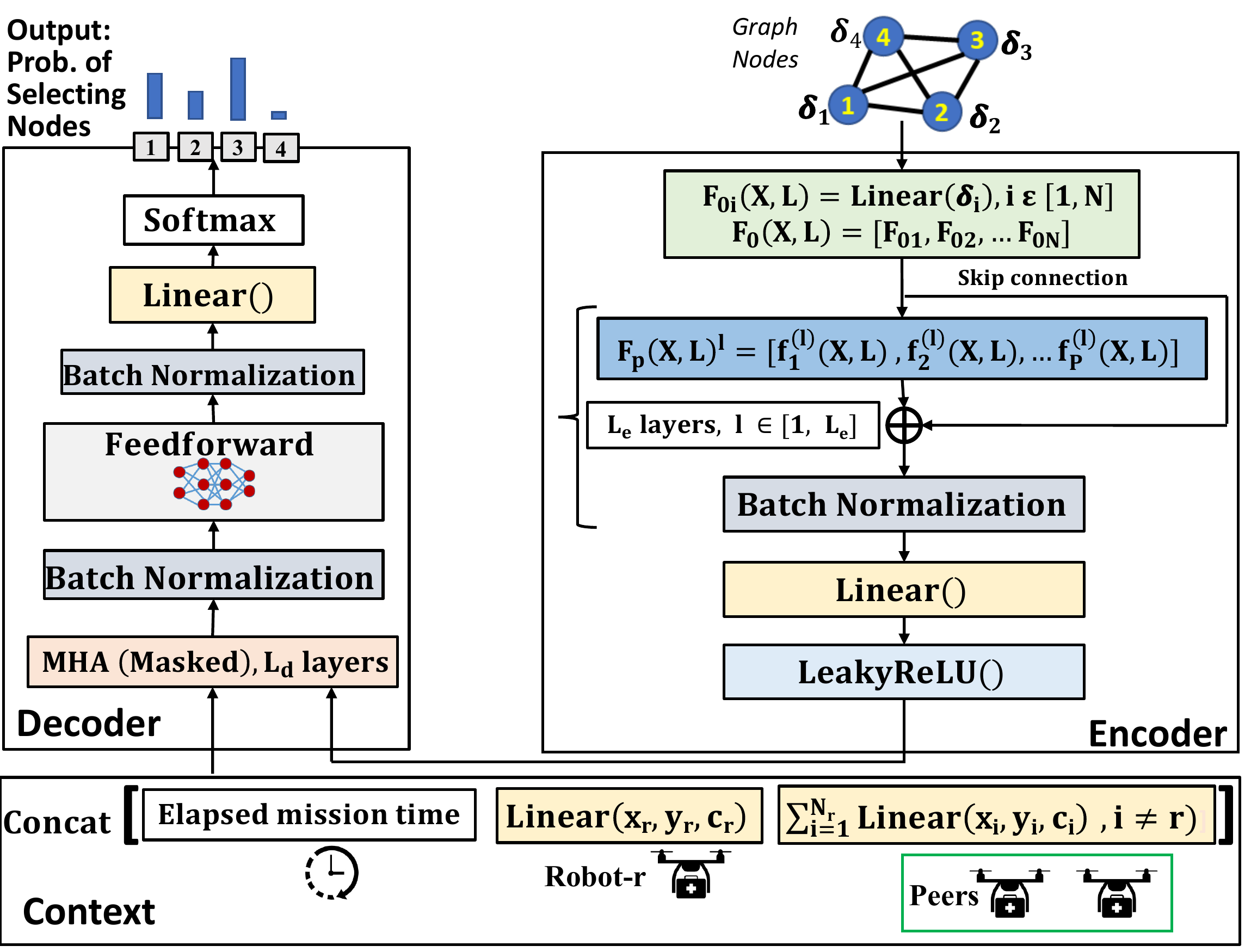}
    \vspace{-0.3cm}
    \caption{CapAM architecture including context, encoder \& decoder.}
    \label{fig:encoder_decoder}
    \vspace{-0.7cm}
\end{figure}

\vspace{-.2cm}
In this work, we implement an RL algorithm on an \textbf{encoder-decoder} architecture to learn the optimal MRTA policies over the graph formulation described earlier. An illustration of this architecture is shown in Fig. \ref{fig:encoder_decoder}. The \textbf{encoder} represents each node as a continuous vector infusing both its own properties as well as its local and global structural properties within the graph representation of the task space. The decoder sequentially computes output probabilities for all the available tasks/nodes using the information from the encoder and the current state (\textbf{context}), by applying the MHA (as shown in left part of Fig. \ref{fig:encoder_decoder}). The sequentially computed output probabilities for each node indicate the relative goodness of selecting that node as the next one to be visited by the concerned robot. A detailed description of the architecture of the encoder and decoder is discussed in the following sections using the representative SR-ST type MRTA problem described in section \ref{sec:ProblemFormulation}. 

\subsection{Graph Capsule Convolutional Network based Encoder}
\label{Sec:Encoder}
The main purpose of the encoder is to represent useful information related to a node/task as a learnable continuous vector or tensor, which can then be used by the learning algorithm. For a graph node, the information includes the properties of the node itself (e.g., the coordinates and time deadline of the node), the local neighborhood information for the node, and also the global structural information. General solutions to CO problems such as CVRP, MTSP and MRTA should be permutation invariant, which means that the order by which each node is numbered should not affect the optimal solution. Hence the node encoding must also be permutation invariant. In this work, we are exploring how a Graph Capsule Convolutional Neural Network (GCAPCN) can be implemented for learning local and global structures with the node properties, with permutation invariant node embedding. GCAPCN is a class of Graph Neural Networks (GNN), introduced in \cite{Verma2018} to address the drawbacks (e.g., permutation invariance) of Graph Convolutional Neural Networks (GCN), and to enable the encoding of global information based on \textbf{capsule networks} presented in \cite{Hinton}. The advantage of GCAPCN lies in capturing more local and global information, compared to conventional aggregation operations used in GNN such as summation or standard convolution operations. 

\subsubsection{Graph Capsule Convolutional Neural Networks}
\label{Sec:GCAPCN}

 Let $X \in \mathbb{R}^{N \times |\delta_{i}|}$, be the node feature matrix, where $|\delta_{i}|$ is the input dimension for each node $i$. The standard graph Laplacian is defined as $L = D - A \in \mathbb{R}^{N \times N}$, where $D$ is the degree matrix and $A$ is the adjacency matrix of the graph. A capsule vector is computed using a Graph Capsule function based on different order of statistical moments, as shown in the equations later in this section.
We first compute a feature vector $F_{0i}$ for each node by linear transformation of the node properties $\delta_{i}$,  as $ F_{0i} = \delta_{i}.W_{0} + b_{0}$ for all $i \in [1,N]$, where
 $W_{0} \  \varepsilon \  \mathbb{R}^{h_{0} \times |\delta_{i}|}, \ \ b_{0} \  \varepsilon \   \mathbb{R}^{h_{0} \times  1}$, and $h_{0}$ is the length of the feature vector. For an MRTA problem with time constraints, $\delta_{i} = [x_{i}, y_{i}, d_{i}]$, where $x_i$, $y_i$, and $d_i$ are the $x$ coordinate, $y$ coordinate, and the time deadline of location/node $i$. Note that this method for encoding is not limited to a specific MRTA problem, and can also be extended or modified for MRTA problems with nodes having other features such as workload.
 Each feature vector $F_{0i}, i \in [1,N]$ is then passed through a series of Graph capsule layers, where the output from the previous layers is used to compute a matrix $f_{p}^{(l)} (X, L)$ using a graph convolutional filter of polynomial form as given by:
\vspace{-.3cm}
\begin{equation}
    \label{Eq:GraphCapsuleFunction}
    f_{p}^{(l)} (X, L) = \sigma(\sum_{k=0}^{K}L^{k}(F_{(l-1)}(X,L)^{\circ p})W_{pk}^{(l)}) \vspace{-.1cm}
\end{equation} \vspace{-.2cm}

\noindent Here $L$ is the graph Laplacian, $p$ is the order of the statistical moment, $K$ is the degree of the convolutional filter, $F_{(l-1)}(X,L)$ is the output from layer $l -1$, $F_{(l-1)}(X,L)^{\circ p}$ represents $p$ times element-wise multiplication of $F_{(l-1)}(X,L)$. 
Here, $F_{(l-1)}(X,L) \in \mathbb{R}^{N \times h_{l-1}p}$, $W_{pk}^{(l)} \in \mathbb{R}^{h_{l-1}p \times h_{l}}$. The variable $f_{p}^{(l)} (X, L) \in \mathbb{R}^{N \times h_{l}}$ is a matrix where each row is an intermediate feature vector for each node $i \in [1,N]$, infusing nodal information from $L_{e} \times K$ hop neighbors, for a value of $p$. The output of layer $l$ is obtained by concatenating all $f_{p}^{(l)} (X, L)$, as given by:
\vspace{-.2cm}
\begin{equation}
    \label{Eq:LayerOutput}
    F_{l}(X,L) = [f_{1}^{(l)} (X, L), f_{2}^{(l)} (X, L), ...f_{P}^{(l)} (X, L)]
\end{equation}
\noindent
Here $P$ is the highest order of statistical moment, and $h_{l}$ is the node embedding length of layer $l$. We consider all the values of $h_{l}$ (where $l \in [0, L_{e}]$) to be the same for this paper.
Equations \ref{Eq:GraphCapsuleFunction} and \ref{Eq:LayerOutput} were computed for $L_{e}$ layers, where each layer uses the output from the previous layer ($F_{l-1}(X,L)$). Adding more layers helps in learning the global structure, however, this can affect the performance by increasing the number of learnable parameters (compared to the size of the problem), leading to over-fitting.
The output from the final layer is then passed through a feed-forward layer so that the final feature vector has the right dimension ($h_l$) to be fed into the decoder as shown in Fig. \ref{fig:encoder_decoder}. 
\vspace{-.2cm}
\subsection{Attention-based Decoder and Context}
\label{Sec:Decoder}
 The main objective of the MHA-based decoder is to use the information from the encoder and the current state as context or query, and thereof choose the best task by calculating the probability value of getting selected for each (task) node. In this case, the first step is to feed the embedding for each node (from the encoder) as \textbf{key-values} ($\mathcal{K}$, $V$), since inputs for MHA are key-value pairs \cite{DBLP:journals/corr/VaswaniSPUJGKP17}. The key $K$ and value $V$ for each node is computed by two separate linear transformations of the node embedding  obtained from the encoder.  The next step is to compute a vector representing the current state, also known as the \textbf{context} (as shown in bottom left of Fig. \ref{fig:encoder_decoder}). 
 In our case studies, the context for the multi-head attention or MHA layer in the decoder consists of the following five features: \textbf{1)} elapsed mission time; \textbf{2)} work capacity of the robot taking decision; \textbf{3)} current location of the robot taking decision; \textbf{4)} current destination of robot's peers; and \textbf{5)} work capacity of peer; all concatenated to a single vector of length $h_{q}$, which then undergoes a linear transformation to get a vector of length $h_{l}$ also called the query $Q$. It is important to note that the \textbf{context} can be modified to account for other features such as remaining ferry-range and payloads of robots, thus promoting the flexibility to apply this architecture to a wider range of MRTA problems than those specifically studied in this paper. 
Figure \ref{fig:encoder_decoder} illustrates the structure of the decoder.

Now the attention mechanism can be described as mapping the query ($Q$) to a set of key-value ($\mathcal{K},V$) pairs. The inputs, which are the query ($Q$) is a vector, while $\mathcal{K}$ and $V$ are matrices of size $h_{l} \times N$ (since there are $N$ nodes). The output is a weighted sum of the values $V$, with the weight vector computed using the compatibility function expressed as:
\vspace{-.2cm}
\begin{align}
   \label{attention_function}
    \text{Attention}(Q,\mathcal{K},V) = \text{softmax}(Q^{T}\mathcal{K}/\sqrt{h_{l}})V^{T} 
     \vspace{-.2cm}
\end{align} 
Here $h_{l}$ is the dimension of the key of any node $i$ ($k_{i} \in \mathcal{K}$).
In this work, we implement a MHA layer in order to determine the compatibility of $Q$ with $\mathcal{K}$ and $V$. The MHA implemented in this work is similar to the decoder implemented in  \cite{Kool2019} and \cite{DBLP:journals/corr/VaswaniSPUJGKP17}.
As shown in \cite{DBLP:journals/corr/VaswaniSPUJGKP17} the MHA layer can be defined as:
\vspace{-.2cm}
\begin{equation}
    \label{mha}
    \text{MHA}(Q,\mathcal{K},V) = \text{Linear}(\text{Concat}(\text{head}_{1}\dots \text{head}_{h_{e}})) 
    \vspace{-.2cm}
\end{equation}

Here $\text{head}_{i} = \text{Attention}(Q,\mathcal{K},V)$ and $h_{e}$ (taken as 8 here) is the number of heads.
The feed-forward layer is implemented to further process the mapping that results from the MHA layer, and transform it to a dimension that is coherent with the number of nodes in the task-graph ($N$). The interjecting batch normalization layers serve to bound values of a specific batch using the mean and variance of the batch. 
The final \texttt{softmax} layer outputs the probability values for all the nodes. Here, the next task to be done is then chosen based on a greedy approach, which means that the node with the highest probability will be chosen. The nodes which are already visited will be masked (by setting their probability as $0$) so that these nodes are not available for selection in the future time steps of the simulation of the multi-robot operation.
\vspace{-.2cm}
\subsection{Training Algorithm}
\label{Sec:LearningAlgorithm} \vspace{-.1cm}

The training algorithm used here is REINFORCE.
Other relevant details for training are shown in Table \ref{table:training_algo_info_MRTA}. This training process can readily be further advanced in the future through the adoption of advanced state-of-the-art policy gradient algorithms such as Proximal Policy Optimization \cite{schulman2017proximal} and the Actor-Critic method \cite{Bahdanau2017}. 
For each epoch of the training, two sets of data (MRTA operation scenarios) are used, which are the training set and the validation set. The training data set is used to train the model ($\theta_\text{CapAM}$) while the validation set is used to update the baseline model ($\theta_\text{CapAM}^{BL}$). Both the training data and validation data generated from the distribution of MRTA scenarios are explained later in section \ref{Sec:Baseline_method_dataset}. 
The size of the training data and the validation data used in this paper is mentioned in Table~\ref{table:training_algo_info_MRTA}. 
\vspace{-.1cm}

\section{Experimental evaluation} 
\label{sec:Experimental_evaluation}

\vspace{-.2cm}
\subsection{Comparison with baselines}
\label{Sec:BaselineComparison}

\subsubsection{Details on dataset for training and testing}
\label{Sec:Baseline_method_dataset}
\begin{wraptable}[10]{R}{4.7cm}
\vspace{-.9cm}
\footnotesize
\caption{Settings for model training in CapAM and AM-RL}
\vspace{-0.7cm}
\label{table:training_algo_info_MRTA}
\vskip 0.15in
\begin{center}
\begin{small}
\begin{sc}
\footnotesize
\begin{tabular}{|l|c|}
\toprule
 Details &  Values \\
\midrule
\textit{Algorithm}    & \textit{REINFORCE}  \\
\textit{Baseline}   & \textit{Rollout}  \\
\textit{Epochs}   & 100\\
\textit{Training samples}   & 500,000  \\
\textit{Validation samples}   & 10,000   \\
\textit{Optimizer}   & \textit{Adam}  \\
\textit{Learning step size} & 0.0001 \\
\textit{Training frequency} & 500 samples  \\
\bottomrule
\end{tabular}
\end{sc}
\end{small}
\end{center}
\vskip -0.1in
\vspace{-0.4cm}
\end{wraptable}

We consider the (NP-hard) MRTA problem with time and capacity constraints also known as Task Allocation Problem with Time and Capacity (TAPTC), as described in \cite{Mitiche2019}.
Each training sample has 100 tasks, and are located randomly within a 100 $\times$ 100 grid map. Each task $i$ has a time deadline $50 \leq d_{i} \leq 600$, and a workload $10 \leq w_{i} \leq 30$. Each sample has $n_r$ number of robots where $2 \leq n_r \leq 7$. The initial positions of the robots in a sample is also chosen randomly within the grids.
Each robot $j$ has a work capacity of $c_{j}$  where $1 \leq c_{j} \leq 3$. All the robots move at a speed of $1$ unit. A task $i$ is considered to be completed only if a robot $j$ visits node $i$ and spends a time of $w_{i}/c_{j}$. All the training samples are generated such that all the associated variables (mentioned above) follow a uniform distribution within their respective bounds. 
The average training time per epoch for CapAM is $~\sim$10 mins.
In order to evaluate and compare the generalizability of CapAM-trained models with that of baselines,
we used the dataset in \cite{Mitiche2019} (found in \cite{dataset}). This dataset, which has a similar distribution of scenarios as used in training CapAM, consists of 96 test cases with 100 tasks and a varying number of robots ($2, 3, 5, 7$ robots), and the speed of every robot is considered to be the same (1 unit/s). The 96 cases are divided into ones with tight and slack deadlines. Each category can be further divided into 4 sub-categories based on the fraction of tasks ($25\%$, $50\%$, $75\%$, and $100\%$) that have normally distributed tasks deadlines. 
For example, $25\%$ indicates that there are 25 tasks with deadlines normally distributed between the limits $d_{\texttt{low}}$ and $d_{\texttt{high}}$, while the remaining 75 tasks have a deadline of $d_{\texttt{high}}$. 
For group 1 (tight deadline), the value of $d_{\texttt{high}}$ is half of that for group 2 (slack deadline). Further description of the test cases can be found in \cite{Mitiche2019}.


\subsubsection{Test for Generalizability}
\label{Sec:TestForGeneralizability}
We compare our method with the following four baseline methods over the stated test cases: 

 \textbf{i)} \textbf{Iterated Local Search (ILS)}: This is an online metaheuristic iterated search algorithm \cite{VANSTEENWEGEN20093281}, where the output of one iteration is partially used as the input to the next iteration. During each iteration, the best solution is improved by a perturbation step, followed by a local search.
 
\textbf{ii)} \textbf{Enhanced Iterated Local Search (EILS)}: EILS is also an online metaheuristic iterated search method \cite{Mitiche2019}, with an improved perturbation step as compared to \cite{VANSTEENWEGEN20093281}.
    
\textbf{iii)} \textbf{Bi-Graph MRTA (BiG-MRTA)}: BiG-MRTA~\cite{ghassemi2019decmrta} is an online method based on the construction and matching of a bipartite graph. In this method, a bipartite graph is constructed to connect robots to tasks, with the weights of connecting edges determined by an incentive model as a function of the tasks' features and robots' states. This decomposes the problem and yield a measure of robot-task pairing suitability. A maximum weighted matching problem is then solved by each robot to identify the optimal task assignments that maximize a net incentive for the team. 
    
\textbf{iv)} \textbf{Attention Mechanism based RL (AM-RL)}: The AM-RL method \cite{Kool2019} consists of an encoder-decoder architecture purely based on attention mechanism.  To implement this method for our problem, it is adapted to a multi-robot setting by making the following changes: 1) The node properties defined in Section~\ref{sec:ProblemFormulation} are used here; 2) The context for the MHA in the decoder is modified to be the same as that in CapAM; and 3) The cost function used for training is changed to that in Eq.~\eqref{eq:objectiveFunction}. We set the parameters which correspond to the best performing model for a CVRP with 100 locations. This AM-RL model is trained using the same sample distribution and settings as respectively described in Sections \ref{Sec:Baseline_method_dataset}. The average training time per epoch is found to be $\sim$11 mins for AM. The architectural difference between AM-RL and CapAM is mainly in the encoder, with their performance comparison intended to show the impact of better encoding for learning over graphs.
\begin{wraptable}[10]{r}{3.6cm}
\vspace{-.7cm}
\footnotesize
\caption{Comparison of computing time to generate entire solution sequence, averaged over 96 cases}
\vspace{-0.75cm}
\label{table:computation_time}
\vskip 0.15in
\begin{center}
\begin{small}
\begin{sc}
\footnotesize
\begin{tabular}{|l|c|}

\toprule
 Method &  Time (sec) \\
\midrule
\textit{ILS}    & ~1.342 \\
\hline
\textit{EILS}   & ~0.992   \\
\hline
\textit{BiG-MRTA}   & ~0.301  \\
\hline
\textit{AM-RL} &  ~0.086  \\
\hline
\textit{CapAM}   & ~0.085  \\

\bottomrule
\end{tabular}
\end{sc}
\end{small}
\end{center}
\vspace{-.1cm}
\end{wraptable}





The computation time to generate the entire sequence of task assignments, as required by the non-learning baselines and the learnt policies of CapAM and AM, are compared in Table \ref{table:computation_time} -- reported as an average over the 96 test cases. The real-time performance advantage of the learnt models, which are $>$10 times faster than the non-learning online methods (ILS/EILS/BiG-MRTA) is readily evident here. 
\begin{figure}[!t]
    \centering
    \includegraphics[width=0.4\textwidth]{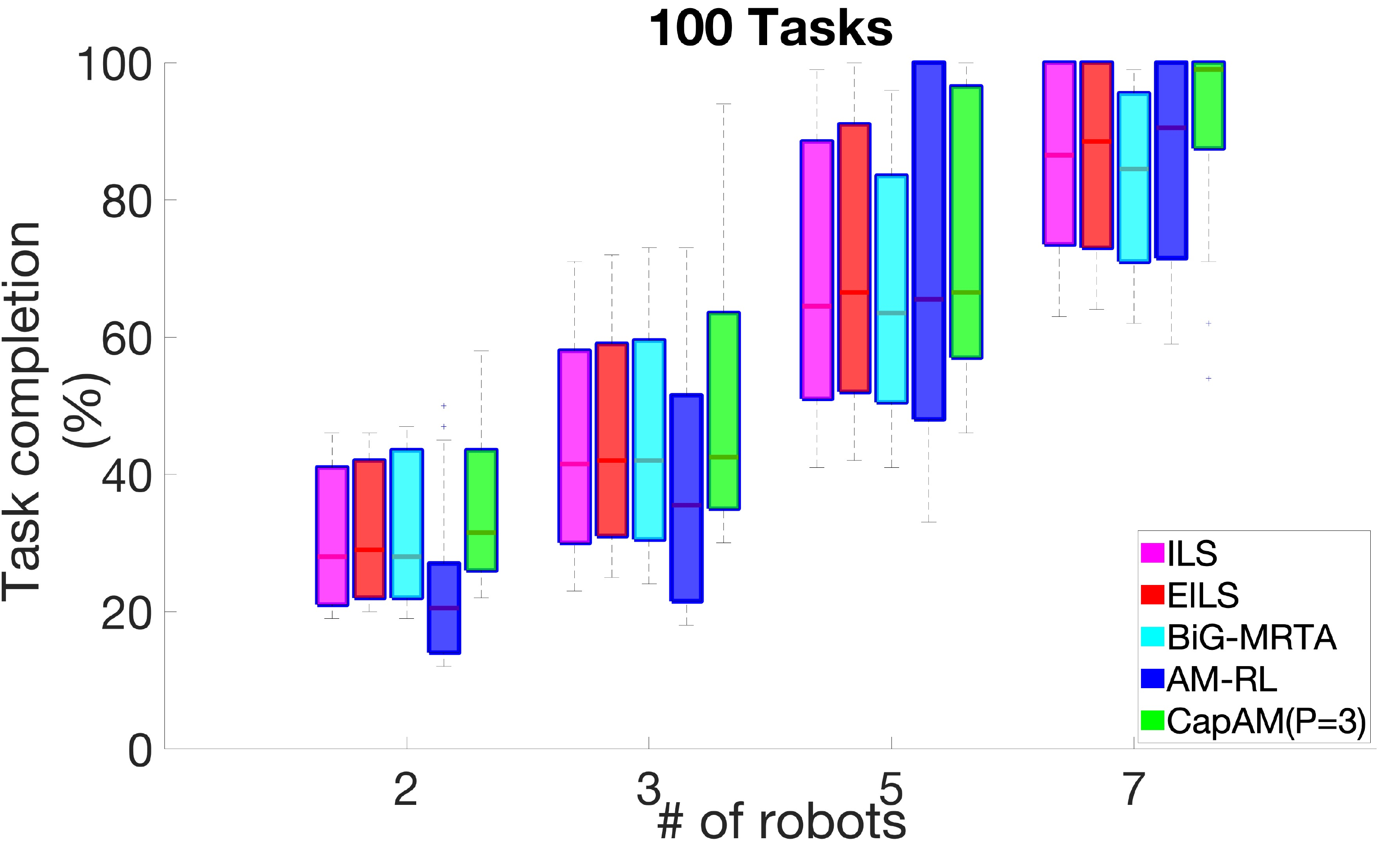}
    \vspace{-0.3cm}
    \caption{Comparison of the performance (on test cases) of CapAM with other baseline methods. Here CapAM has $K$ = 2, $P$ = 3, and $L_{e}$ = 1.}
    \label{fig:gini_test_cases}
    \vspace{-0.7cm}
\end{figure}
Figure \ref{fig:gini_test_cases} shows the box plots for task completion performance over the 96 test cases divided into 4 categories (each with 24 cases) based on the number of robots (2,3,5, and 7). As seen from Fig. \ref{fig:gini_test_cases}, CapAM obtains a better performance compared to the baseline methods. To provide statistical evidence of the advantage of CapAM over the baseline methods, we performed a pairwise T-test of CapAM \textit{vs} each baseline on the 24 cases in each category, with results summarized in Table \ref{table:ttest_pvals}; here, the null hypothesis is that the difference between the values of the two sets has a mean equal to $0$. Considering a 5$\%$ significance level, Table \ref{table:ttest_pvals} shows that the \textbf{p-value} from the T-test is generally less than $0.05$, which indicates the rejection of the null hypothesis -- CapAM's performance on the test cases is thus significantly better than of the baselines.

\begin{table}[H]
\vspace{-.4cm}
\footnotesize
\centering
\caption{\textbf{p-value} for pairwise (CapAM vs. baselines) T-Test corresponding to Fig. \ref{fig:gini_test_cases}.} \vspace{-.3cm}
\label{table:ttest_pvals}
\begin{tabular}{|c|c|c|c|c|}
\hline
\multirow{2}{*}{\textbf{\# robots}} & \multicolumn{4}{c|}{\textbf{CapAM vs}}                            \\ \cline{2-5} 
                                    & \textbf{ILS} & \textbf{EILS} & \textbf{BiG-MRTA} & \textbf{AM-RL} \\ \hline
\textbf{2}                          & 2.03e-4      & 0.002         & 0.003             & 1.6e-11        \\ \hline
\textbf{3}                          & 0.0307       & 0.079         & 0.038             & 2.4e-8         \\ \hline
\textbf{5}                          & 0.024        & 0.163         & 8.8e-4            & 0.42           \\ \hline
\textbf{7}                          & 0.025        & 0.049         & 0.003             & 0.031          \\ \hline
\end{tabular}%
\vspace{-.4cm}
\end{table}

\subsection{Parametric and Scalability Analysis}
\label{Sec:Parametric-ScalabilityAnalysis}
\vspace{-.2cm}
Here, parametric analyses is performed w.r.t. the key parameters of $P$ and $L_{e}$ in the encoder of CapAM. Higher values of $P$ promotes greater information infusion for a given neighborhood; with increasing $L_{e}$ or $K$ (and fixed $P$), information from (great number of) $L_{e} \times K$ hop neighbours is considered in the node embeddings. Increasing $K$, $P$ and $L_{e}$ however increases the number of learnable parameters, which can lead to over-fitting and deteriorating performance for unseen cases. 
Our numerical experiments with higher values of $K$ and $L_{e}$ corroborated this expectation, and hence we fixed $K=2$ and $L_{e}=1$. 
Hence this section focuses on the impact of $P$, particularly in generalizing, and scaling to larger-sized problems beyond training -- both of which are premised as major advantages of learning local structural information.
To this end, we train 3 different models on the 100-task problem, with varying $P$ from 1 to 3, and fixed $K$ and $L_{e}$.
Then the CapAM (and for comparison AM-RL) models that are trained for 100 tasks and 2-7 robots are implemented on two sets of unseen problems -- one with the same size as in training, and another with numbers of tasks and robots scaled up by a factor of 2 (training/testing tasks are drawn from the same distribution). 

Table \ref{table:CAM_P_var} shows the effect of $P$ on scalability of CapAM, which is also compared to AM-RL. Results are reported in terms of task completion $\%$ averaged over 24 samples in each category (based on \# tasks and robots). It can be seen that all CapAM models with any $P$ perform significantly better than AM-RL. For the problems of the same size as training (100-task cases), the task completion \% gap between CapAM and AM-RL is between $3\%$ and $13.5\%$. For the problems of double the size as training (200-task cases), the task completion \% gap between CapAM and AM-RL is between $1.9\%$ and $17.6\%$. In both cases, the gap (CapAM vs. AM-RL) is the highest when task-to-robot ratios are high. 
These observations directly demonstrate the effectiveness of the novel encoder used in CapAM, which is designed to capture local structural information much better than in attention mechanisms. 
On comparing the CapAM models with different $P$ (in Table \ref{table:CAM_P_var}), it can be seen that ones with $P = 1$ perform slightly poorer compared to corresponding ones with $P=2$ and $P=3$, with the  difference between the latter two being marginal. Thus increasing $P$ beyond 2 may not provide further performance improvements. 
\begin{table}[H]
\vspace{-.35cm}
\footnotesize
\caption{Performance of CapAM (in terms of \% of task completed) with $P=1,2,3$ for unseen problem sets with the same and twice the number of tasks and robots as in training.}
\vspace{-.3cm}
\label{table:CAM_P_var}
\resizebox{.48\textwidth}{!}{%
\footnotesize
\begin{tabular}{|c|c|c|c|c|c|}
\hline
\textbf{\# tasks} &
  \multicolumn{1}{l|}{\textbf{\# robots}} &
  \multicolumn{1}{l|}{\textbf{CapAM($P$=3)}} &
  \multicolumn{1}{l|}{\textbf{CapAM($P$=2)}} &
  \multicolumn{1}{l|}{\textbf{CapAM($P$=1)}} &
  \multicolumn{1}{l|}{\textbf{AM-RL}} \\ \hline
\multirow{4}{*}{\textbf{100}} & \textbf{2}  & 34.6 & \textbf{36.2} & 32.6 & 22.7 \\ \cline{2-6} 
                              & \textbf{3}  & \textbf{51.1} & 50.3 & 49.4 & 38.0 \\ \cline{2-6} 
                              & \textbf{5}  & \textbf{74.9} & \textbf{74.9} & 73.2 & 70.2 \\ \cline{2-6} 
                              & \textbf{7}  & \textbf{94.8} & 94.1 & 91.2 & 85.6 \\ \hline
\multirow{4}{*}{\textbf{200}} & \textbf{4}  & \textbf{28.2} & 27.1 & 24.4 & 10.7 \\ \cline{2-6} 
                              & \textbf{6}  & 46.7 & \textbf{47.0} & 45.2 & 29.4 \\ \cline{2-6} 
                              & \textbf{10} & 67.8 & \textbf{68.6} & 66.5 & 61.4 \\ \cline{2-6} 
                              & \textbf{14} & \textbf{77.8} & 77.5 & 76.2 & 74.3 \\ \hline
\end{tabular}%
}
\vspace{-.4cm}
\end{table}

\footnote{The codes for this paper can be found in \url{https://github.com/adamslab-ub/CapAM-MRTA}}.
\section{Conclusion}
\label{sec:Conclusion}\vspace{-.2cm}
In this paper, we proposed a new graph neural network architecture called CapAM, to learn policies for MRTA problems involving tasks with time deadlines and robots with work capacity. This new architecture incorporates an encoder based on capsule networks and a decoder based on the attention mechanism, along with a context module to feed the state of the robots and the environment. To learn the features of the encoder and decoder, the problem has been posed as an RL problem and solved using the policy gradient algorithm, REINFORCE. In addition, the proposed architecture is found to provide effective MRTA policies over varying task size and robot-team size. The new CapAM architecture demonstrated better performance compared to other state-of-the-art baselines, which include both a learning based method (AM-RL) and non-learning based methods (ILS, EILS, BiG-MRTA).
The CapAM architecture with its capsule based node embedding showed that learning local and global structural information of the task graph results in better generalizability over unseen test cases, as observed from comparing its performance with AM-RL (that uses a different node embedding). The computational cost analysis showed that the trained CapAM model takes only a few milliseconds to yield a task-assignment decision, making it highly suited for operations that require time-sensitive online decisions. The advantage of using local neighborhood information was also empirically evident from the scalability analysis on the defined MRTA problem, where CapAM demonstrated superior performance when applied to graphs with a larger number of tasks/nodes and robots in the team. 

\textbf{Future directions:} Firstly, computational analysis of training feasibility when episodes must be simulated over more realistic virtual robotics environment remains a open direction of research across all learning-based approaches. Further, to enable transition of these MRTA methods to application, in the future we should consider dynamic tasks, environment uncertainties, and partially observable state spaces within the CapAM architecture. 
\bibliographystyle{./bibliography/IEEEtran}
\bibliography{./bibliography/IEEEabrv,./bibliography/IEEE_ICRA}

\end{document}